\documentclass[showpacs,twocolumn]{revtex4}
\ifx\pdfoutput\undefined
\usepackage{graphicx}
\else
\usepackage{graphicx}
\fi

\usepackage[center]{subfigure}
\usepackage{bm}
\usepackage{amsmath,amssymb}
\usepackage{latexsym}
\usepackage{natbib}

\def\beq{\begin{equation}}
\def\eeq{\end{equation}}
\def\nn{\nonumber}

\begin{document}

\title{Fluctuation-driven Turing patterns}
\author{Thomas Butler\footnote{Present Address: Department of Physics and Department of Chemical Engineering, Massachusetts Institute of Technology,
77 Massachusetts Avenue, Cambridge MA, 02139}}
\author{Nigel Goldenfeld}
\affiliation{Department of Physics and Institute for Genomic Biology,
University of Illinois at Urbana Champaign, 1110 West Green Street, Urbana, IL 61801 USA}

\date{\today}

\begin{abstract}

Models of diffusion driven pattern formation that rely on the Turing mechanism are utilized in many areas of science.  However, many such models suffer from the defect of requiring fine tuning of parameters or an unrealistic separation of scales in the diffusivities of the constituents of the system in order to predict the formation of spatial patterns.  In the context of a very generic model of ecological pattern formation, we show that the inclusion of intrinsic noise in Turing models leads to the formation of ``quasi-patterns" that form in generic regions of parameter space and are experimentally distinguishable from standard Turing patterns.  The existence of quasi-patterns removes the need for unphysical fine tuning or separation of scales in the application of Turing models to real systems.

\end{abstract}



\maketitle

The study of the emergent spatiotemporal patterns in physical or biological systems is an exciting and fruitful line of research in physics and in many other disciplines such as chemistry, ecology, animal biology, and neuroscience \cite{CROS93,LEVI92,TURI53,BRES02,KOCH94}.  Examples include patterns on animal coats \cite{MURR81}, engineered bacterial systems \cite{LU10}, chemical pattern formation \cite{CAST90}, mussel population densities \cite{KOPP08}, and Rayleigh-Benard convection in fluids \cite{RAYL16}. 

One particularly satisfying aspect of these studies is that insight into the origins of one kind of pattern often yields insight into the origins of patterns in entirely different systems.  A key example is the Turing mechanism \cite{TURI53}.  Turing's argument, which will be described in detail below, showed how diffusion, which is typically thought of as a randomizing influence, can give rise to spatial pattern formation when there are two or more classes of degrees of freedom (species) with ``activator" and ``inhibitor" dynamics.  This mechanism has been proposed as an explanation for an enormous variety of systems including short ($<10$m) length scale patchiness in planktonic ecosystems \cite{LEVI76,MALC08,DAVI92,ABRA98}, patterning in plant-resource systems \cite{REIT08}, patchiness in insect abundance \cite{MARO97}, stripe and spot patterns on the coats of animals \cite{MURR81}, patterns in mussel beds \cite{KOPP08} and even the geometric visual hallucinations experienced by shamans and users of hallucinogenic drugs \cite{BRES02,BUTL10}.  

However, in spite of the seeming success of the Turing mechanism in explaining patterns across many disciplines, the partial differential equations representing the dynamics of systems with Turing patterns typically require unphysical fine tuning of parameters or separation of scales in the diffusivities of the different species in order to predict pattern formation \cite{CAST90,LEVI76,MIMU78,WILS99, BAUR07,BUTL09b,KOCH94}.  The requirement that the system either have fine tuning of kinetic parameters or a separation of scales in diffusivities in order to predict patterns, is unphysical for many applications.  Both the of these issues will be referred to below as the ``fine tuning problem," even though fine tuning is only strictly needed when a separation of scales in diffusivities is not present.  To resolve the fine tuning problem for Turing patterns we show that a full statistical mechanical treatment of Turing patterns, where fluctuations due to the discrete nature of the degrees of freedom in the system -- intrinsic noise -- are included, the fine tuning problem is resolved \cite{BUTL09b}.  

It may seem counterintuitive to claim that including fluctuations resolves the fine tuning problem for Turing patterns because fluctuations are generally expected to \textit{destabilize} ordered states such as spatial patterns.  This is the rule in standard statistical mechanics \cite{GOLD92} and many statistical mechanical models in ecology \cite{KATO98,DAVI08}.  However, exceptions exist in systems out of equilibrium.  For example, careful experiments on Rayleigh-Benard convection have shown that fluctuations can drive the formation of convection rolls in fluid dynamics that would not form in the absence of fluctuations \cite{WU95}.  In ecology, recent theoretical work and careful data analysis have shown that the observed cyclic population dynamics of predator-prey systems can be explained in many cases by fluctuation driven cycles in time \cite{MCKAN05,LUGO08,BUTL09,PINE07}.  Similar phenomena have been predicted in evolutionary game theory and systems biology \cite{BLAD10,MCKAN07}.  In cell biology, simulating the interactions of individual proteins in discrete time and space in a model of proteins that regulate cell division in e-coli results in pattern formation over a wider range of parameters than the corresponding reaction-diffusion partial differential equations \cite{HOWA03}. Thus it seems possible that a full many body treatment of the Turing mechanism that incorporates intrinsic noise will resolve the fine tuning problem.

The purpose of this paper is to present an analyisis of the Turing mechanism with intrinsic noise included to resolve the fine tuning problem.  The analysis results in a derivation of a phase diagram and to power spectra with experimentally distinctive and relevant properties.  This paper is an expansion and elaboration of our paper \cite{BUTL09b} which originally reported the resolution of the fine tuning problem of Turing instabilities through the incorporation of intrinsic noise.  We will first review the Turing mechanism, and then present an extremely simple model of the Turing mechanism for planktonic predator-prey populations that we then analyze in detail.  The results of the analysis show that in large regions of parameter space predicted by deterministic modeling to have only trivial spatial states a new kind of spatial pattern emerges that we call a ``quasi-pattern."  The quasi-pattern state is analogous to intrinsic noise driven ``quasi-cycles" recently discovered in the time domain \cite{MCKAN05}.  Quasi-patterns are recognizable immediately as spatial patterns, but with a few important, experimentally relevant, differences from patterns predicted with deterministic analysis.  The final sections of the paper will focus on possible experimental tests and extensions of the theory developed in the body of the paper.  We focus on a model of planktonic predator-prey interactions throughout the paper for simplicity and also because predator-prey systems have been extensively analyzed theoretically \cite{LEVI76,MIMU78, BAUR07,MALC08, MOBI06} and there is beginning to be an experimental literature  \cite{MARO97,WILS99}.  However, we emphasize that the goal of this paper is insight into the general interactions of intrinsic fluctuations with the Turing mechanism for pattern formation and that the results should be valid for most models of Turing instabilities.  Evidence for this assertion is provided by the recent replication of our results on the Brusselator model of chemical pattern formation \cite{BIAN10}, which additionally pointed out that the phase boundary can differ for different species of reactants, a model of embryonic pattern formation \cite{SCOT10}, as well as our own forthcoming results on pattern formation on the visual cortex \cite{BUTL10}.

\section{The Turing mechanism}

The Turing mechanism in its most basic form requires two different species that react and diffuse.  One species, the ``activator," diffuses relatively slowly, and catalyzes (activates) both its own production and the production of the second species.  The second species, the ``inhibitor," diffuses faster, and reduces (inhibits) the concentration of both the activator species and itself.  These combined mechanisms lead to pattern formation from random initial conditions.  We illustrate the mechanism with the example of predator-prey dynamics with random initial conditions
\\
\noindent 1. Random regions of activator (prey) with higher local concentrations reproduce rapidly, leading to dense clumps of activator species that then begin to diffuse.
\\
\noindent 2. Rapidly diffusing inhibitors (predators) are produced in the neighborhood of the high density autocatalyzing clumps of prey.
\\
\noindent 3. The predators inhibit the spread of the prey clumps through their production in the neighborhood of prey clumps.  The autoinhibitory nature of predators prevents them from overwhelming the prey population.   

\begin{figure}[ht]
\begin{center}
\includegraphics[width=3.5in]{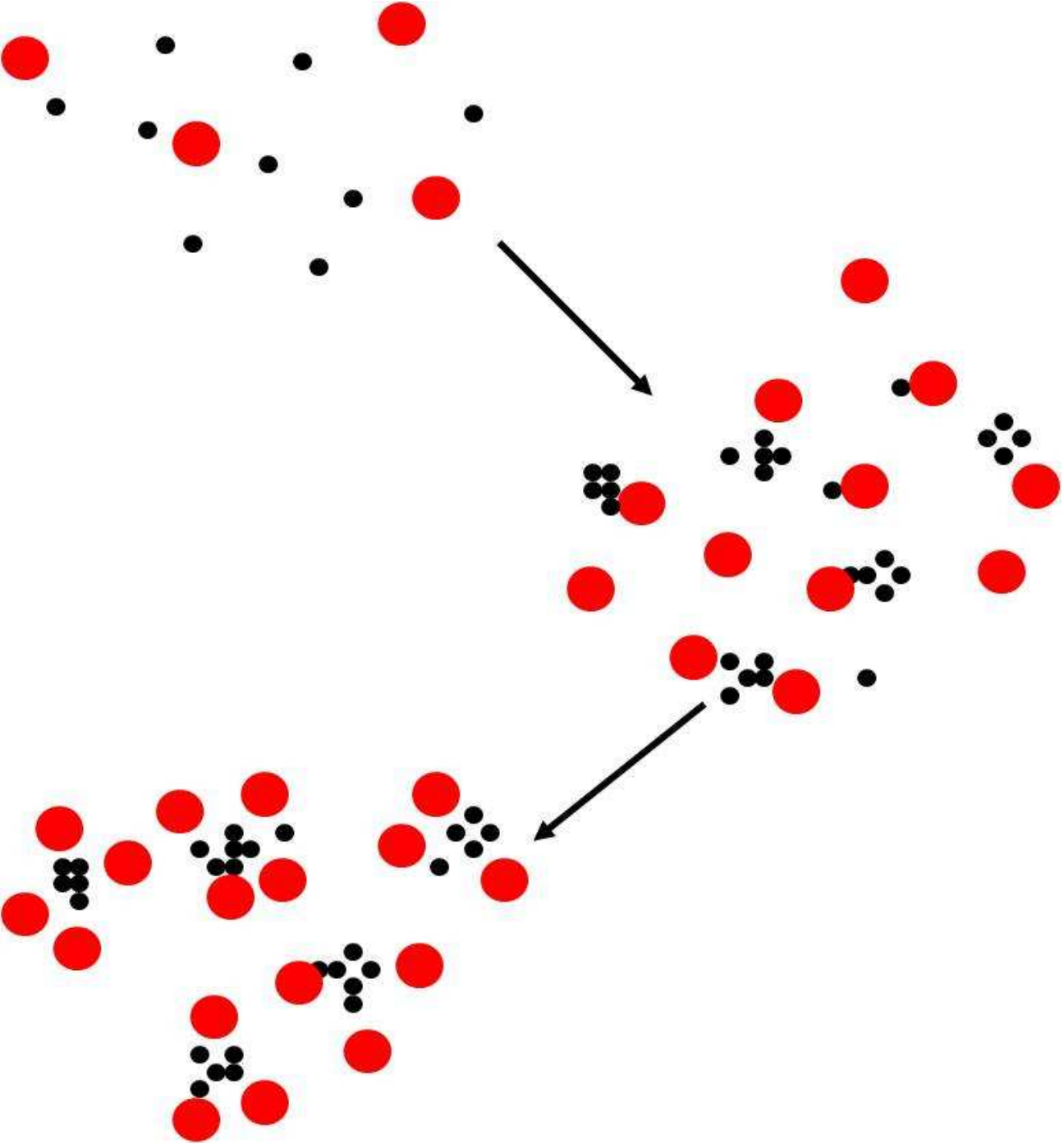}
\caption{Illustration of the steps of the Turing mechanism as described in the text.  The figure should be viewed from top to bottom.  The prey (activators) are represented by black dots, and the predators (inhibitors) are represented by red dots.}
\label{ch5:fig1a}
\end{center}
\end{figure}

These steps, summarized in fig. \ref{ch5:fig1a}, show how activator-inhibitor dynamics can lead to spontaneous pattern formation \cite{TURI53}.  As was noted above, formalizing this argument into standard deterministic reaction-diffusion equations results in models that only exhibit
Turing patterns if the predator (inhibitor) diffusivity is much larger than
the prey (activator) diffusivity or the parameters are fine tuned
\cite{LEVI76,MIMU78,WILS99, BAUR07,CAST90,TURI53}.  Note that consistent with the existence of pattern forming systems which do not apparently display very large
separation of diffusivities \cite{MARO97,REIT08} the qualitative argument made above for pattern formation does not depend on very large differences in diffusivities, nor on additional kinetic details.  

\section{Turing patterns in the Levin-Segel model}

One of the simplest models of Turing patterns is drawn from ecological pattern formation and was
originally introduced to model plankton-herbivore
dynamics \cite{LEVI76}.  The reaction diffusion equations for this model are
\begin{equation}
\begin{split}
\partial_t\psi &=\mu\nabla^2\psi+b\psi+e\psi^2-(p_1+p_2)\psi\varphi \\
\partial_t\varphi &= \nu\nabla^2\varphi+p_2\varphi\psi-d\varphi^2
\end{split}
\label{ch5:1}
\end{equation}
\noindent where the plankton population $\psi$ is the activator, as can be seen by the nonlinear growth term $e\psi^2$, and the herbivore population $\varphi$ is the inhibitor due to the predation terms $p\psi\varphi$ and the competition term $-d\varphi^2$.  The nonlinear growth term $e\psi^2$ was origingally introduced to be a proxy for predator satiation \cite{LEVI76} but can also be interpreted as an Allee effect, wherein many species have enhanced reproduction at higher concentrations (for a review, see \cite{COUR99}).  

Setting $p_1=0$ and $p_2=p$, the model contains a stable homogeneous coexistence state when
\begin{equation}
\begin{split}
p>e \ \text{and} \ p^2>de
\end{split}
\label{ch5:1a}
\end{equation}
\noindent with stationary fixed point populations given by
\begin{align}
\psi_s=\frac{b d}{p^2-de}, \ \ \varphi_s=\frac{b p}{p^2-de}
\label{ch5:1b}
\end{align}
\noindent It contains a Turing instability if \cite{LEVI76}
\begin{align}
\frac{\nu}{\mu}>\left(\frac{1}{\left(\sqrt{p/d}-\sqrt{p/d-e/p}\right)}\right)^2
\label{ch5:1c}
\end{align}
The model is only valid when the coexistence fixed point is stable.  Outside of that regime, a plankton regulation term such as $-f\psi^3$ is required to make the model valid.  For the present analysis, we assume that $f\psi^3$ is sufficiently small to be ignored.  The fine tuning problem can be illustrated in this model by taking the set of $O(1)$ kinetic
parameters $b=1/2, \ e=1/2, \ d=1/2$ and $p=1$.  With these
parameters Eq. \ref{ch5:1c} shows that non-generic diffusivities,
$\nu/\mu>27.8$, are required for pattern formation.  Similar results
are obtained for other generic parameter sets.

\subsection{Extrinsic noise driven pattern formation}
To gain preliminary insight into the effects of intrinsic noise on the Levin-Segel model, we can analyze the effects of unconserved extrinsic noise on the linearized dynamics of the Levin-Segel model as in previous studies of extrinsic noise driven pattern formation \cite{GARC93,CARR04,SEIB07,BUTL09b}.  While not identical, expansion schemes such as the system size expansion \cite{VANK92} indicate that the effects of extrinsic noise and intrinsic noise on the linearized dynamics of reaction diffusion systems are closely related.  Additionally, we will use the calculation of the effects of extrinsic noise on the Levin-Segel model to predict observable differences between unconserved extrinsic and intrinsic noise driven pattern formation.

To calculate the effects of extrinsic noise, we look at the
Fourier transformed dynamics of the fluctuations from the coexistence
fixed point with added white noise $\xi$, variance $C$.  These dynamics
are given by
\begin{equation}
-i\omega \bm{x}=\bm{A}\bm{x}+\xi
\label{ch5:1d}
\end{equation}

\noindent The matrix $\bm{A}$ is the Fourier transformed stability
matrix and $\bm{x}$ is the vector of deviations from equilibrium of predator and prey populations respectively,
\begin{equation}
\bm{A} = \left( \begin{array}{cc} -\nu k^2 -p\psi_s& p \varphi_s \\
                                                                    -p\psi_s               & -\mu k^2+e\psi_s
                                                                    \end{array} \right)
\label{ch5:1e}
\end{equation}

\noindent Simple manipulations yield the average power spectrum
\begin{equation}
\begin{split}
&P(k,\omega)= C\left[p^2 \varphi_s^2+(e\psi_s-\mu k^2)^2 \right] \times \bigg[\big(p b\psi_s+\mu\nu k^4-\omega^2 \\ &-\psi_s k^2 e\nu\left(1-\frac{p\mu}{e\nu}\right)\big)^2 +\omega^2((e-p)\psi_s-(\mu+\nu)k^2)\bigg]^{-2}
\end{split}
\label{ch5:1f}
\end{equation}

To a crude approximation, Eq. \ref{ch5:1f} predicts that patterns
(indicated by peaks in the power spectrum) form whenever $e\nu>p\mu$, and that without noise and away from a classical Turing instability the power spectrum is zero.  As anticipated, the condition $e\nu>p\mu$ can be satisfied easily and avoids the fine tuning problem.  However, the calculation with the extrinsic noise considered here differs in important ways from the intrinsic noise case, such as the determination of the strength of the noise and the presence of diffusive noise.  As will be shown below, these differences lead to experimentally distinguishable differences in the resulting spatiotemporal patterns.

\section{Predator-prey model with intrinsic noise}

To systematically include the effects of intrinsic noise requires a model defined at the level of individual organisms, since intrinsic noise is generated by the stochastic nature of individual birth and death events as well as the stochastic interactions between individual organisms.  Such a description of the dynamics at the individual level is called an individual level model (ILM).  One simple way to define an ILM is to specify the reactions that can take place in a well mixed patch of volume $V$.  To include space, a lattice of patches can be considered with additional reactions corresponding to movement of predator and prey organisms between the patches.  With parameters to specify the relative rates of the reactions, a model of individual level interactions on a single patch that incorporates intrinsic noise is fully specified.  

For an ILM version of the Levin-Segel model we consider the following
reactions
\begin{align}
P  &\stackrel{b}{\rightarrow}PP \nn \\
PP &\stackrel{e/V}{\rightarrow} PPP \nn \\
PH &\stackrel{p_1/V}{\rightarrow} H \nn \\
PH &\stackrel{p_2/V}{\rightarrow}HH \nn \\
HH &\stackrel{d/V}{\rightarrow} H
\label{ch5:2}
\end{align}

\noindent where $P$ denotes plankton and $H$ denotes herbivores, with
the parameters as described above. Stochastic trajectories of $H$ and
$P$, enumerated by $m$ and $n$ respectively, are described by the
master equation
\begin{align}
&\partial_t P(m,n) = b(-nP(m,n)+(n-1)P(m,n-1)) \nn \\
&+\frac{e}{V}((n-1)(n-2)P(m,n-1) - n(n-1)P(m,n)) \nn \\
&+\frac{p_1}{V}(-mnP(m,n)+(m)(n+1)P(m,n+1)) \nn \\
&+\frac{p_2}{V}(-mnP(m,n)+(m-1)(n+1)P(m-1,n+1)) \nn \\
&+\frac{d}{V}\left[(m+1)mP(m+1,n)-m(m-1)P(m,n)\right]
\label{ch5:3}
\end{align}

The master equation, which is exactly equivalent to the specification of the model as a collection of reactions in Eq. \ref{ch5:2}, can then be used to analyze the ILM version of the Levin-Segel model by applying techniques from non-equilibrium statistical mechanics.  

\subsection{Field theory representation of the model}
While several options exist for analysis of the master equation, such as direct expansion of the master equation \cite{VANK92}, we analyze the master equation by a mapping to field theory, because it is convenient for handling spatially extended systems.  To analyze the master equation using the techniques of field theory, we introduce the operators
\begin{align}   
a|m,n\rangle &= m|m-1,n \rangle \nn \\
\hat{a}|m,n\rangle &= |m+1,n\rangle \nn \\
\left[ a, \hat{a} \right] &= 1	\nn \\
c|m,n\rangle &= n|m,n-1 \rangle \nn \\
\hat{c}|m,n\rangle &= |m,n+1\rangle \nn \\
\left[ c, \hat{c} \right] &= 1 \:\:
\label{ch5t:4}
\end{align}

\noindent and the state $|\psi\rangle = \sum P(n)|n\rangle$.  These definitions allow the master equation to be mapped to a bosonic field theory \cite{DOI76, GOLD84, MIKH81, PELI85, JANS08}.  
As an explicit example of how to convert the master equation to a field theory, consider the master equation corresponding to the second reaction in Eq. \ref{ch5:2} alone. 
\begin{equation}
\partial_t P(n)  = \frac{e}{V}[(n-1)(n-2)P(n-1) - n(n-1)P(n)]
\label{ch5t:2}
\end{equation}

\noindent Ignoring $V$ for now, we multiply both sides by $|n\rangle$ and sum over $n$
\begin{align}
&\sum_{n}\partial_t P(n)|n\rangle = \nn \\
&e \sum_{n}\left[(n-1)(n-2)P(n-1)-n(n-1)P(n)\right]|n\rangle
\label{ch5t:3}
\end{align}
 
We next shift the sums, and manipulate the first term in the sum

Let $n'=n-1 \rightarrow n = n'+1$.
\begin{align}
& e\sum_{n'}n'(n'-1)P(n')|n'+1\rangle \;,n' \rightarrow n \nn \\
&= e \hat{c}^3 c^2 \sum P(n)|n\rangle \nn \\
&= e \hat{c}^3 c^2 |\psi \rangle \:\: 
\label{ch5t:5}
\end{align}

We now work out the second term in the sum
\begin{align}
e\sum_{n}n(n-1)P(n)|n\rangle \nn \\
= e \hat{c}\hat{c} c c |\psi\rangle \:\:
\label{ch5t:6}
\end{align}

This yields
\begin{equation}
\partial_t |\psi\rangle = e \left[\hat{c}^3-\hat{c}^2\right]c^2 |\psi \rangle
\label{ch5t:7}
\end{equation}

Similar analyses lead to second quantized forms for the rest of the master equation.  We can now assemble the entire Hamiltonian.  We start by writing the master equation in second quantized form
\begin{widetext}
\begin{align}
\partial_t |\psi\rangle = \left[b(\hat{c}^2-\hat{c})c+\frac{e}{V}(\hat{c}^3-\hat{c}^2)c^2+\frac{p_1}{V}(\hat{a}ac-\hat{a}a\hat{c}c)+  
\frac{p_2}{V}(\hat{a}^2ac-\hat{a}a\hat{c}c)+\frac{d}{V}(1-\hat{a})\hat{a}a^2\right]|\psi\rangle
\label{ch5t:12}
\end{align}
\end{widetext}

\noindent Since the standard definition of the Hamiltonian is
\begin{equation}
\partial_t |\psi\rangle = -\hat{H}|\psi\rangle
\label{ch5t:13}
\end{equation}

\noindent we have
\begin{align}
-\hat{H}&= b(\hat{c}^2-\hat{c})c+\frac{e}{V}(\hat{c}^3-\hat{c}^2)c^2+\frac{p_1}{V}(\hat{a}ac-\hat{a}a\hat{c}c) \nn \\
&+\frac{p_2}{V}(\hat{a}^2ac-\hat{a}a\hat{c}c)+\frac{d}{V}(1-\hat{a})\hat{a}a^2
\label{ch5t:14}
\end{align}

According to the standard mapping using coherent states between Hamiltonians represented by bosonic operators and functional integral representations of the same dynamics with Lagrangians, we can write down the Lagrangian, generalized to space.  As in quantum mechanics, the mapping can be worked out for general Hamiltonians \cite{CARD98, PELI85}.  To generalize to space, we implement a random walk between patches of volume $V$ for every organism as a reaction with rate $\tau_i$ where $i$ is an index for species.  Appropriately rescaled \cite{BUTL09}, the continuum limit and mapping to the functional integral formulation yields the Lagrangian 
\begin{align}  
\mathcal{L} = \hat{a}\partial_t a + \hat{c} \partial_t c - \nu \hat{a} \nabla^2 a - \mu \hat{c}\nabla^2 c \nn \\
+ H (\hat{c}, \hat{a}, c, a)
\label{ch5t:15}
\end{align}

In the Lagrangian formulation of Eq. \ref{ch5t:15}, $\hat{a}$, $\hat{b}$ and their conjugate variables are no longer operators, but are functions that are integrated over as in standard bosonic functional integrals.  The starred variables loosely correspond to noise and the unstarred to values of predator and prey, but direct physical interpretation is not trivial \cite{CARD96,MATT98}.  
The initial conditions are ignored, because the focus of this paper is the long time limit and there is only one attractor in the system.

To transform to more physical variables, the standard Cole-Hopf transformation can be applied to transform the field variables to direct number and noise representations.  This transformation is given by
\begin{align}
a&=ze^{-\hat{z}} \nn \\
\hat{a} &= e^{\hat{z}} \\
c &= \rho e^{-\hat{\rho}} \nn \\
\hat{c} &= e^{\hat{\rho}}
\label{ch5t:16}
\end{align}

\noindent the new field variables $z$ and $\rho$ can be heuristically interpreted as the number of predator and prey respectively (the precise interpretation is that their expectation values correspond, i.e. $\langle f(\rho,z)\rangle=\langle f(N_P,N_H)\rangle$) and the auxiliary fields denoted by carets generate the intrinsic noise, as will be seen below by showing that the minimum of the action, which corresponds to mean field theory is at $\hat{\rho}=\hat{z}=0$. The Lagrangian in the new variables is
\begin{align}
\mathcal{L} &= \hat{x}\partial_t z + \hat{\rho} \partial_t \rho -\nu\hat{z}\nabla^2 z - \mu \hat{\rho}\nabla^2 \rho \nn \\ &-\nu z(\nabla \hat{z})^2 -
\mu\rho(\nabla \hat{\rho})^2+b\rho(1-e^{\hat{\rho}})\nn \\ 
&+\frac{e}{V}\rho^2(1-e^{\hat{\rho}}) + \frac{p_1}{V}z\rho(1-e^{-\hat{\rho}}) \nn \\ &+\frac{c}{V}z\rho(1-e^{\hat{z}-\hat{\rho}})+\frac{d}{V}z^2(1-e^{-\hat{z}})
\label{ch5t:19}
\end{align}  
\subsection{System size expansion}
We now can carry out the system size expansion in the field theoretic formalism.  Other than notation, it is identical to the direct expansion of the master equation reviewed in \cite{VANK92}.  We expand the fields as
\begin{align}
\hat{z} \rightarrow \frac{\hat{z}}{\sqrt{V}}, \; \;
\hat{\rho} \rightarrow \frac{\hat{\rho}}{\sqrt{V}} \nn \\
z=V\varphi +\sqrt{V}\eta, \;\;
\rho = V \psi + \sqrt{V}\xi
\label{ch5t:20}
\end{align}

To perform this expansion to consistent order, it is necessary to expand the exponentials out to second order.  This is because the expansion will promote second order terms to first order.  The result is an expansion of the Lagrangian in the form
\begin{equation}
\mathcal{L}=\sqrt{V}\mathcal{L}_1+\mathcal{L}_2 +O(1/\sqrt{V})
\label{ch5t:20a}
\end{equation}  

We once again carry out the expansion explicitly for the term coupled by $e/V$.  
\begin{align}
&\frac{e}{V}\rho^2(1-e^{\hat{\rho}}) \nn \\
&=\frac{e}{V}\left(V\psi+\sqrt{V}\xi\right)\left(V\psi+\sqrt{V}\xi\right)\left(1-\left(1+\frac{\hat{\rho}}{\sqrt{V}}+\frac{\hat{\rho}^2}{2V}\right)\right) \nn \\
&=-e\left(\sqrt{V}\psi^2\hat{\rho}+\frac{\psi^2\hat{\rho}^2}{2}+2\xi\hat{\rho}\right)+O(1/\sqrt{V})
\label{ch5t:22}
\end{align}

Collecting terms of leading order, $\sqrt{V}$, we have
\begin{align}
\mathcal{L}_1&=\hat{\rho}\partial_t\psi+\hat{z}\partial_t\varphi-\nu\hat{z}\nabla^2\varphi-\mu\hat{\rho}\nabla^2\psi-b\psi\hat{\rho}-e\psi^2\hat{\rho} \nn \\ &+b\varphi\psi\hat{\rho}-c\varphi\psi(\hat{z}-\hat{\rho})+d\varphi^2\hat{z}
\label{ch5t:26}
\end{align}

It is trivial to extract the mean field PDE's by using the Euler-Lagrange equations.  The equations that result are
\begin{align}
\frac{\delta \mathcal{L}_1}{\delta \hat{z}}=\partial_t\psi-\mu\nabla^2\psi+b\psi+e\psi^2-(p_1+p_2)\psi\varphi=0
\label{ch5t:27}
\end{align}
\noindent which is the first of the equations for the Levin-Segel model.  The second equation is
\begin{align}
\frac{\delta \mathcal{L}_1}{\delta \hat{\rho}}=\partial_t\varphi - \nu\nabla^2\varphi+p_2\varphi\psi-d\varphi^2 = 0
\label{ch5t:28}
\end{align}

\noindent again reproducing the Levin-Segel model equation of motion. Note that the auxiliary fields have zero expectation value at mean field, which confirms the interpretation that they correspond to noise.  Now $\mathcal{L}_2$ can be assembled.  The terms in $\mathcal{L}_2$ that are linear in $\eta$ or $\xi$ correspond to the stability matrix of the MFT.  Terms that are quadratic in the hatted variables $\hat{\rho}$ and $\hat{z}$ are noise terms and will be considered next.

Proceeding, we have
\begin{align}
\mathcal{L}_2&=\hat{z}\partial_t\eta+ \hat{\rho}\partial_t\xi-\hat{z}\nu\nabla^2\eta-\hat{\rho}\mu\nabla^2\xi +p_1\eta\psi\hat{\rho}\nn \\ &-p_2\eta\psi(\hat{z}-\hat{\rho})+2d\eta\varphi\hat{z}+b\xi\hat{\rho}+2e\xi\psi\hat{\rho}+b\xi\varphi\hat{\rho} \nn \\&-p_2\xi\varphi(\hat{z}-\hat{\rho})
\label{ch5t:29}
\end{align}

\noindent We convert this into a Fourier transformed matrix form that includes time and space
\begin{equation}
\mathcal{L}_2 = \bm{y}^T\partial_t \bm{x} - \bm{y}^T\bm{A}\bm{x}-\frac{1}{2}\bm{y}^T\bm{B}\bm{y}
\label{ch5t:30}
\end{equation}

\noindent with vectors given by
\begin{align}
\bm{x} = \left( \begin{array}{c} \eta \\   
				 	\xi   \end{array} \right),\				 					     			
\bm{y}=	\left( \begin{array}{c} \hat{z} \\
				 				\hat{\rho}   \end{array} \right)			
\label{ch5t:31}
\end{align}

\noindent so that the predator variables are on top.  The matrix A is the Jacobian of the MFT with space and is given by
\begin{align}
\bm{A} = \left( \begin{smallmatrix} -\nu k^2 +p_2\psi-2d\varphi & p_2 \varphi \\
																	-(p_1+p_2)\psi				 & -\mu k^2+b+2e\psi-(p_1+p_2)\varphi 
																	\end{smallmatrix} \right)
\label{ch5t:32}
\end{align}

The matrix for the correlations of the noise is given by 
\begin{align}
\bm{B}=\left( \begin{smallmatrix} d\varphi^2+p_2\varphi\psi + \nu\varphi k^2& -p_2 \varphi\psi \\
																	-p_2 \varphi\psi 				 & b\psi+e\psi^2+b\varphi\psi+p_2\varphi\psi +\mu\psi k^2
																	\end{smallmatrix} \right)
\label{ch5t:34}
\end{align}

We also now note that $\mathcal{L}_2$ is in the form of a Lagrangian in the
Martin-Siggia-Rose (MSR) response function formalism for Langevin
equations \cite{MART73, JANS76}. 

\subsection{The power spectrum}

We now extract the stochastic differential equations (SDE) that govern the dynamics of the fluctuations, and calculate the power spectrum of the fluctuations.  The Langevin equations from the response function formalism are
\begin{align}
i\omega\bm{x}=\bm{A}\bm{x}+\bm{\gamma}(\omega) \nn \\
\langle \gamma_i (\omega) \gamma_j(-\omega)\rangle = B_{ij}
\label{ch5t:35}
\end{align}

\noindent We solve formally to obtain
\begin{align}
\bm{x} = (\bm{A} + i \omega)^{-1}\bm{\gamma}(\omega) \equiv \bm{D}(\omega)^{-1}\bm{\gamma}(\omega) 
\label{ch5t:36}
\end{align}

\noindent The power spectrum is
\begin{align}
\langle x_1 x_1^* \rangle &= \frac{\langle(D_{22}\gamma_1-D_{12}\gamma_2)(D_{22}^*\gamma_1-D_{12}^*\gamma_2)\rangle}{|det(D)|^2} \nn \\
&= \frac{|D_{22}|^2 B_{11}-2D_{12}Re(D_{22})B_{21}+|D_{12}|^2 B_{22}}{|det(D)|^2}
\label{ch5t:37}
\end{align}

To find the phase diagram, take $p_1\rightarrow 0$, $p_2=p$.  This simplification does not substantially change the dynamics of the model.  In terms of elements of the stability matrix $\bm{J}\equiv \bm{A}(k=0,\omega=0)$, the denominator of the power spectrum is  
\begin{align}
\det(D)&=(J_{11}+i\omega-\nu k^2)(J_{22}+i\omega-\mu k^2)-J_{12}J_{21} \nn \\
&= det(J) + i\omega (Tr(J)-(\mu+\nu)k^2) \nn \\&-(J_{11}\mu+J_{22}\nu)k^2 +\mu\nu k^4-\omega^2
\label{ch5t:38}
\end{align}

The full expression for the power spectrum is
\begin{widetext}
\begin{align}
P(k,\omega)=\frac{|D_{22}|^2 B_{11}-2D_{12}Re(D_{22})B_{21}+|D_{12}|^2 B_{22}}{(det(J)+\mu\nu k^4-\omega^2-(J_{11}\mu k^2+J_{22}\nu k^2))^2+\omega^2(Tr(J)-(\mu+\nu)k^2)^2}
\label{ch5t:39}
\end{align}
\end{widetext}
Recall the fixed point values at coexistence are
\begin{align}
\varphi=\frac{pb}{p^2-de} \nn \\
\psi = \frac{db}{p^2-de}  \nn \\
\label{ch5t:42}
\end{align}
\noindent Using the fixed point values, the matrix $\bm{A}$ can be further simplified to 
\begin{align}
\bm{A} = \left( \begin{array}{cc} -\nu k^2 -p\psi& p \varphi \\
																	-p\psi				 & -\mu k^2+e\psi 
																	\end{array} \right)
\label{ch5t:45}
\end{align}
Now we evaluate the determinant of the ODE stability matrix ($\bf{J}$ above) and the trace
\begin{align}
\det(\bm{J})=p\psi b 
\label{ch5t:46}
\end{align}
The trace is 
\begin{align}
Tr(\bm{J})=(e-p)\psi
\label{ch5t:47}
\end{align}

Simplifying the denominator in Eq \ref{ch5t:39} yields
\begin{widetext}
\begin{align}
|det(D)|^2=(det(J)+\mu\nu k^4-\omega^2-(J_{11}\mu k^2+J_{22}\nu k^2))^2+\omega^2(Tr(J)-(\mu+\nu)k^2)^2 \nn \\
=(p b\psi+\mu\nu k^4-\omega^2-\psi(-p\mu k^2+e\nu k^2))^2+\omega^2((e-p)\psi-(\mu+\nu)k^2)^2 \nn \\
=\left(p b\psi+\mu\nu k^4-\omega^2-\psi k^2 e\nu\left(1-\frac{p\mu}{e\nu}\right)\right)^2+\omega^2((e-p)\psi-(\mu+\nu)k^2)^2
\label{ch5t:48}
\end{align}
\end{widetext}

The form of the denominator for $\omega=0$ is $(A-Bk^2+Ck^4)^2$, which has a minimum at non zero $k$.  This minimum corresponds to an emergent length scale, and is the first indication of pattern formation.  Systematic demonstration of the emergence of pattern formation requires accounting for the $k$ dependence in the numerator. The noise matrix $\bm{B}$ can be simplified to  
\begin{align}
\bm{B}=\left( \begin{array}{cc} 2p\varphi\psi + \nu\varphi k^2& -p \varphi\psi \\
																	-p \varphi\psi 				 & 2p\varphi\psi +\mu\psi k^2
																	\end{array} \right)
\label{ch5t:51}
\end{align}

Notice the symmetry in the noise correlations.  We now can expand out the numerator of Eq. \ref{ch5t:39}
\begin{widetext}
\begin{align}
&|D_{22}|^2 B_{11}-2D_{12}Re(D_{22})B_{21}+|D_{12}|^2 B_{22} \nn \\
&=|e\psi-\mu k^2+i\omega|^2(2p\varphi\psi + \nu\varphi k^2)+2p\varphi(e\psi-\mu k^2)(p\varphi\psi)+p^2\varphi^2(2p\varphi\psi+\mu\psi k^2) \nn \\&=(e\psi-\mu k^2)^2 (2p\varphi\psi + \nu\varphi k^2) +\omega^2 (2p\varphi\psi + \nu\varphi k^2)+2p^2\varphi^2\psi(e\psi-\mu k^2)+p^2\varphi^2(2p\varphi\psi+\mu\psi k^2)
\label{ch5t:52}
\end{align}
\end{widetext}

This gives the final form of the power spectrum
\begin{widetext}
\begin{align}
P(k,\omega)=\frac{(e\psi-\mu k^2)^2 (2p\varphi\psi + \nu\varphi k^2) +\omega^2 (2p\varphi\psi + \nu\varphi k^2)+2p^2\varphi^2\psi(e\psi-\mu k^2)+p^2\varphi^2(2p\varphi\psi+\mu\psi k^2)}{\left(pb\psi+\mu\nu k^4-\omega^2-\psi k^2 e\nu\left(1-\frac{p\mu}{e\nu}\right)\right)^2+\omega^2((e-p)\psi-(\mu+\nu)k^2)^2}
\label{ch5t:53}
\end{align}
\end{widetext}

\section{Analysis of the power spectrum}
\subsection{Phase diagram for quasi-patterns}

The expression for the power spectrum in Eq. \ref{ch5t:53} is not very illuminating, and does not simplify a great deal.  To find quasi-patterns we note that the highest power of $k$ in the denominator of Eq. \ref{ch5t:53} is larger than the highest power in the numerator.  That means for sufficiently large $k$, the power spectrum is always decreasing.  Thus, to show the existence of a maximum, it is sufficient to show that for small $k$, the power spectrum is increasing.  This can be shown by computing $\frac{dP}{dk^2}$ and evaluating at $k^2=0$.  When this expression is greater than 0, there is pattern formation.  This yields the analytical criterion
\begin{align}
\frac{\nu}{\mu}>\frac{p^3(5p^2 + 7de)}{e(4p^4 + 5p^2 de + 3 d^2 e^2)}
\label{ch5t:55}
\end{align}

\noindent This criterion is much less stringent than the criterion for Turing instabilities.
The conditions for a Turing instability are
\begin{align}
\frac{\nu}{\mu}>\left( \frac{1}{\sqrt{p/d}-\sqrt{p/d-e/p}} \right)^2
\label{ch5t:56}
\end{align}

\begin{figure}[ht]
\begin{center}
\includegraphics[width=3.5in]{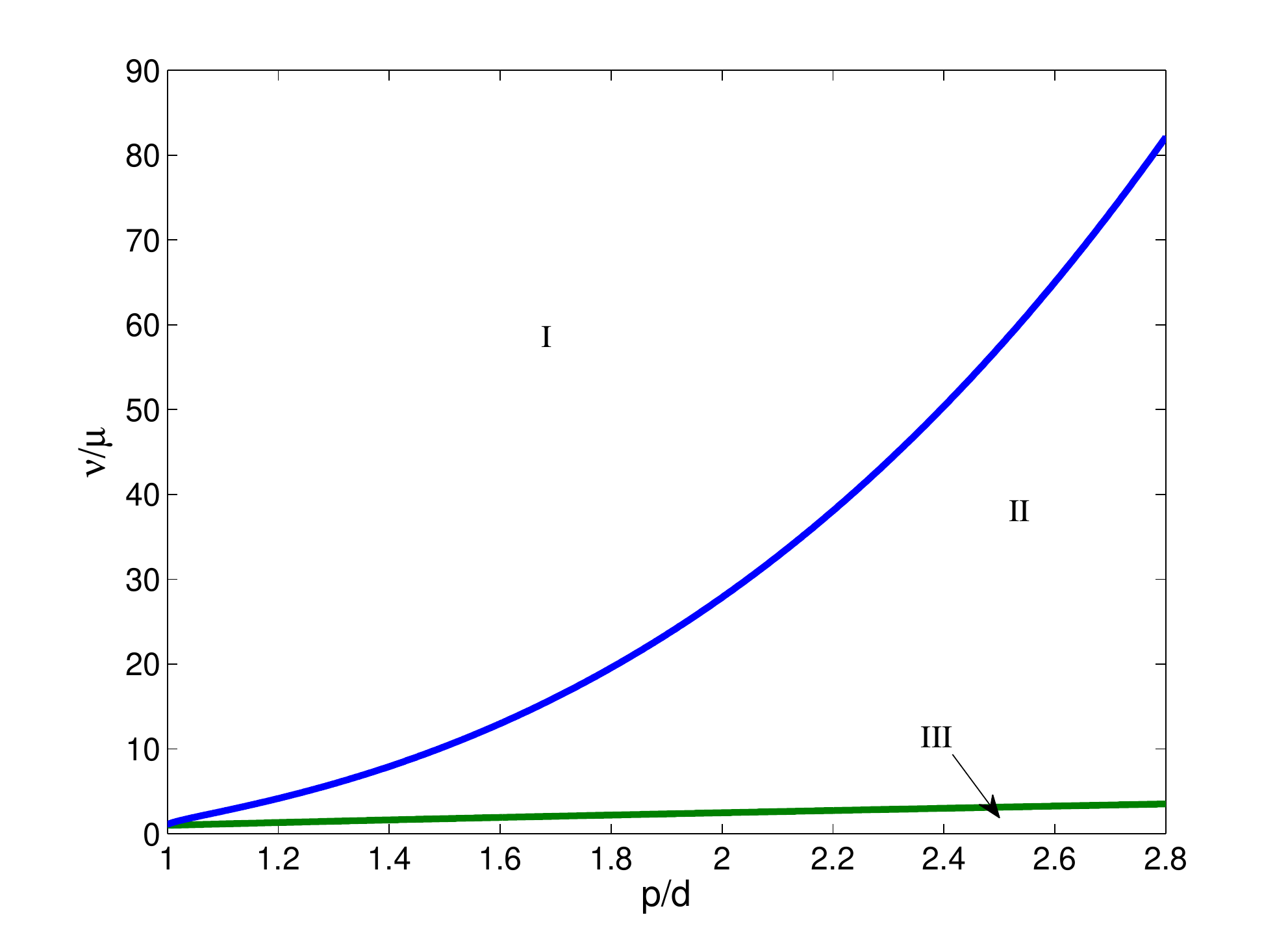}
\caption{Phase diagram over stable parameter region in $p/d$.  Region I is MFT level pattern formation, region II contains fluctuation driven quasi-patterns, and region III is a spatially homogeneous phase.}
\label{ch5:fig1}
\end{center}
\end{figure}

For the generic parameters $b=1/2,\; p=1,\; d=1/2,\; e=1/2$ the criterion \ref{ch5t:55} yields $\nu/\mu>2.48$, while the Turing condition yields $\nu/\mu>27.8$.  This behavior is typical of generic parameters.  The phase diagram of the system bears out this conclusion as shown in figure \ref{ch5:fig1}.

\begin{figure}[ht]
\begin{center}
\includegraphics[width=3.5in]{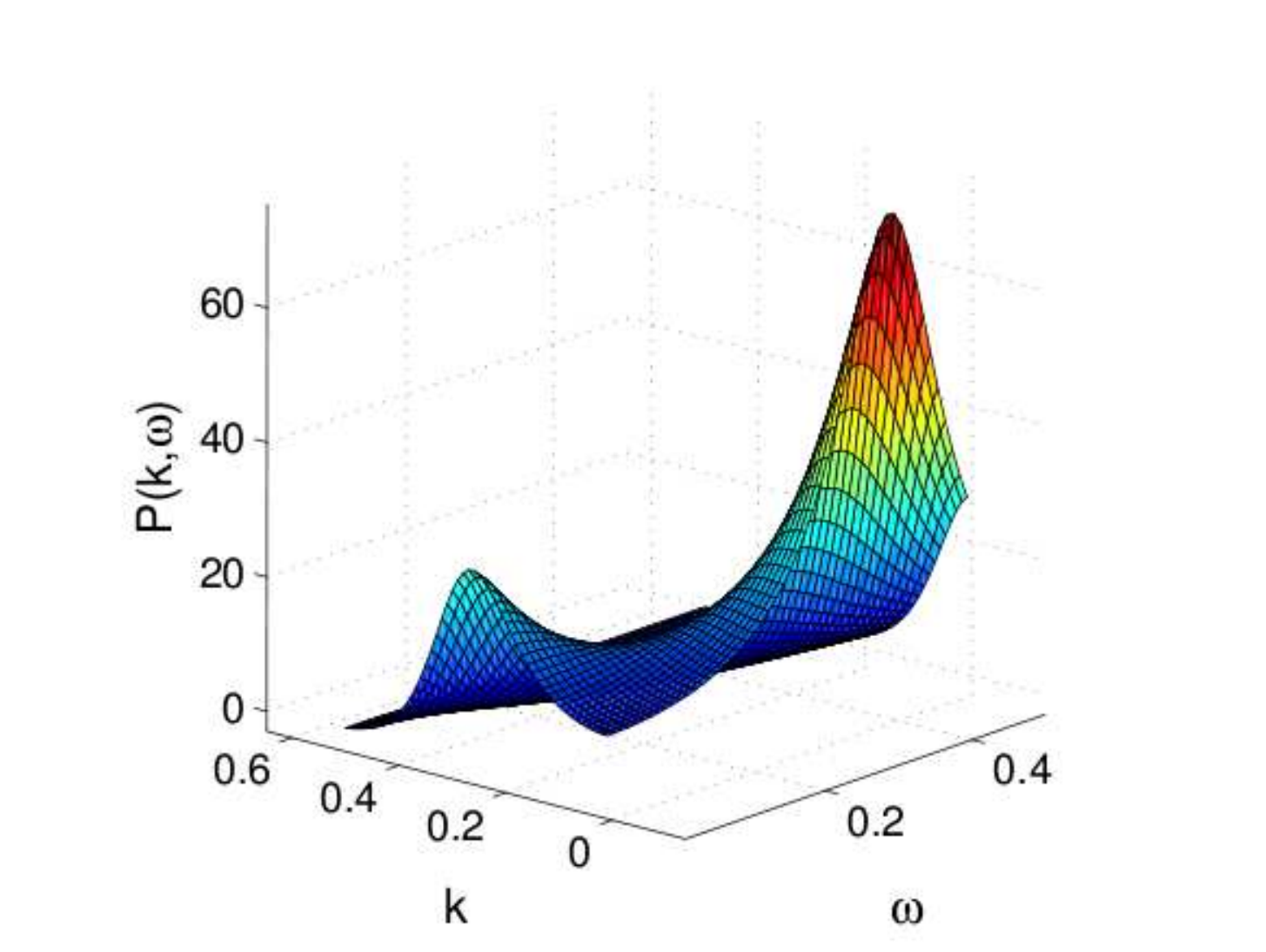}
\caption{Power spectrum with p=1, $\nu/\mu$=15}
\label{ch5:fig2}
\end{center}
\end{figure}

An additional feature of the model is that oscillations and spatial
pattern formation are essentially decoupled.  This means that the model
predicts global population oscillations and spatial pattern formation,
but not traveling waves.  The mathematical origin of this can be seen
in Eq. \ref{ch5:1f}.  The $k^2$ term with a negative coefficient at
$\omega=0$ is quickly overwhelmed by the positive $k^2$ dependence of
the $\omega^2$ term as the frequency begins to grow.  In the power
spectrum (fig. \ref{ch5:fig2}) this can be seen as the deep valley between
the peaks in $k$ and $\omega$.  This interpretation is supported by
preliminary simulations.

\subsection{Wavelength of fluctuation driven patterns}

To a fairly good approximation, the wavelength of the Turing quasi-patterns can be calculated.  The wavelength corresponds to the wave vector that maximizes the power spectrum.  To calculate that value, consider the numerator of the power spectrum only at $\omega=0$.
\begin{equation}
\left(pb\psi+\mu\nu k^4-\psi k^2 e\nu\left(1-\frac{p\mu}{e\nu}\right)\right)^2)^2
\label{ch5:57}
\end{equation}

The minimum of this expression will correspond with reasonable accuracy to the real wavelength and can be obtained through straightforward calculation to be
\begin{equation}
\lambda_m=\frac{2 \pi}{k_m}=\sqrt{\frac{2 \mu}{\psi}\left(1-\frac{c\mu}{e\nu}\right)}
\label{ch5:58}
\end{equation}

This shows that for a fixed ratio of diffusivities, the wavelength increases as the square root of the diffusivity.  In addition, while the phase diagram of the system (fig \ref{ch5:fig1}) and therefore the presence of Turing quasi-patterns depends on diffusivity only through the ratio $\nu/\mu$, the wavelength of the patterns depends on the values of the diffusivities.

This calculation also implies that the wavelength of the quasi-patterns is closely related to the wavelength of patterns in the region of the phase diagram where patterns are generated at mean field.  In the standard theory of Turing patterns, patterns are formed when the homogeneous steady state is unstable to perturbations with a specific set of wave vectors $k$.  The wavelength is then the wavelength corresponding to the mode that is most unstable.  In the calculation above, we have picked out the mode that in mean field theory corresponds to the slowest decaying mode as the wavelength of the quasi-patterns.  This is because the denominator of the power spectrum is equal to the product of the eigenvalues of the stability matrix squared.  This product is smallest for the slowest decaying mode, which is also the mode that will go unstable in mean field theory first as parameters are varied.  Therefore the wavelength of the quasi-patterns corresponds to the wavelength of the mean field patterns.  

\subsection{Period of quasi-cycles}

A similar calculation to the calculation above for the wavelength of the quasi-patterns can be carried out for the period of the quasi-cycles.  Consider the denominator of the power spectrum with $k=0$  
\begin{equation}
\left(p b \psi-\omega^2\right)^2+\omega^2((e-p)\psi)^2
\label{ch5:59}
\end{equation}

Analogous to the wavelength calculation, we seek the minimum in $\omega$.  Simple calculation yields a period of
\begin{equation}
T=\frac{2\pi}{\omega_m}=\frac{4\pi}{\sqrt{2bp\psi-(e-p)^2\psi^2}}
\label{ch5:60}
\end{equation}

Similar arguments to those for the wavelength indicate that the period for the quasi-cycles is approximately the period for the stable spirals present in mean field theory \cite{MCKAN05}.

\section{Distinguishing quasi-patterns and quasi-cycles from other spatiotemporal patterns}

To distinguish spatiotemporal patterns generated by intrinsic noise from those generated by feedbacks alone (i.e. mean field patterns) or by extrinsic noise, it is necessary to develop theoretical predictions that differ for each of these cases.  Previous work has focused primarily on time series data, focusing on problems such as distinguishing quasi-cycles from limit cycles \cite{PINE07} as well as the task of simply determining the amount of extrinsic versus intrinsic noise in ecosystems \cite{BJOR01}.  This work has confirmed that both extrinsic noise and intrinsic noise are important in real ecosystems for populations such as temperate songbirds in Norway, and the beetle species Tribolium \cite{BJOR01,DENN88,SAET00} and that quasi-cycles are present in real ecological time series data\cite{PINE07}. The work also confirms that the importance of intrinsic noise decreases as population density increases, in line with the expectation that the scale of intrinsic noise depends on the scale of the population density \cite{SAET00}. 

While separate signatures of quasi-cycles and quasi-patterns will be discussed below, one common feature that distinguishes quasi-cycles and quasi-patterns from their counterparts in mean field theory is that they depend on the concentration of the population being studied.  To leading order only the fluctuations have patterns, implying that the local populations can be written as mean value plus fluctuations scaled by the size of a locally well mixed region (see below).  Thus the amplitude of the patterns relative to the mean population size of the fluctuation driven patterns will change as the size of a locally well mixed area changes, while the relative amplitude of mean field patterns and limit cycles would not change.  Such a variation of the size of a locally well mixed area could presumably be used to detect quasi-patterns and quasi-cycles.

\subsection{Distinguishing quasi-cycles from limit cycles}

Given a population that has oscillatory abundance in time, theory indicates that the oscillations can come from either quasi-cycles driven by noise or from population density dependent feedbacks alone, perturbed by noise (mean field cycles).  The key difference mathematically is that the power spectrum of limit cycles has a pole at its frequency while the power spectrum of quasi-cycles does not.  In the time domain, this means that the cycles driven by intrinsic noise have a short correlation time while limit cycles have an infinite correlation time.  Since poles do not exist in real population data due to stochasticity, finite size populations, measurement error, etc. what this means for real data is that there is a separation of scales between the correlation time of limit cycles and quasi-cycles.  This was first pointed out in detail by Pineda-Krch et al. \cite{PINE07}.  These authors also showed that wolverine population cycles are likely quasi-cycles, while the celebrated lynx-hare cycles from the Hudson Bay company's trapping records are most likely limit cycles \cite{PINE07}.

Other studies of the role of intrinsic noise have focused on intrinsic noise contributions compared to extrinsic noise contributions as a function of local population size \cite{BJOR01,SAET00}.  In frequency space, the best frequencies to analyze to distinguish the relative importance of noise are high frequencies, corresponding to the short timescale fluctuations of the system.  To extract predictions for the case of intrinsic noise, we look at the large $\omega$ asymptotics of the power spectrum Eq. \ref{ch5t:53} at $k=0$ 
\begin{equation}
P(k=0,\omega)= \frac{2p\psi\varphi}{\omega^2}, \; \; \omega \gg \omega_m
\label{ch5:61}
\end{equation}

\noindent where $\omega_m$ is the modal frequency of the quasi-cycles.  For cycles driven by extrinsic additive noise, we look at the same asymptotics for the power spectrum from the heuristic calculation, which, as we noted above, can be considered as a calculation for extrinsic noise.  In this case, the asymptotic form is
\begin{equation}
P(k=0,\omega)= \frac{p^2\psi^2+e^2\varphi^2}{\omega^{4}}\langle \xi \xi\rangle, \; \; \omega \gg \omega_m
\label{ch5:62}
\end{equation}

\noindent where the variance $\langle \xi \xi\rangle$ is independent of population density and $\omega_m$ is the frequency of the quasi-cycles.  While in this case, both the expressions depend on the square of population density, the decay in $\omega$ has a power of two for intrinsic noise, and of four in the case of extrinsic noise. Thus the tails can be easily distinguished in real data.  

\subsection{Distinguishing quasi-patterns from mean field patterns}

Similar considerations can be applied to quasi-patterns.  While further study is needed, the finite peak in the power spectrum for quasi-patterns indicates that quasi-patterns generically have a shorter correlation length than mean field patterns, which have a pole in their power spectrum at the wavelength of the pattern.  Thus the techniques outlined above for distinguishing mean field limit cycles from quasi-cycles and applied to real populations in \cite{PINE07} translate directly into the space domain from the time domain.  

For distinguishing unconserved extrinsic noise and intrinsic noise contributions, the asymptotics for short wavelength fluctuations can again be derived for the intrinsic and extrinsic noise cases.  For intrinsic noise, we have
\begin{equation}
P(k,\omega=0)=k^{-2}\frac{\varphi}{\nu}, \; \; k \gg k_m
\label{ch5:63}
\end{equation}

\noindent where $k_m$ is the wave vector of the mode of the power spectrum.  For extrinsic noise, we have
 \begin{equation}
P(k,\omega=0)=\frac{k^{-4}}{\nu^2}\langle \xi \xi\rangle, \; \; k \gg k_m
\label{ch5:64}
\end{equation}
 
\noindent where the variance $\langle \xi \xi\rangle$ is independent of population density.  Like the quasi-cycle case, the scaling in $k$ differs by a power of two between the extrinsic and intrinsic noise cases.  Contrary to quasi-cycles in the previous section, the extrinsic and intrinsic noise lead to different powers of population density for large $k$.  This provides a useful tool for distinguishing between the effects of unconserved extrinsic anoise nd intrinsic noise on the formation of patterns especially if the density of the populations can be varied through comparative study of field data in different ecosystems, or through experiments.  These considerations are quite broad, and should qualitatively apply to other systems, such as chemical reaction systems in which quasi-patterns or cycles may be present, such as the Brusselator model of chemical pattern formation \cite{BIAN10}.

Another possibility beyond those considered here is noise manifested through stochasticity in the kinetic parameters of the system as is common in ecological models \cite{MAY73}.  Systematic study of such a model is well beyond the scope of this paper, and is an interesting subject for future research.  However, a simple model of the effects of weak parameter noise on the linearized dynamics shows that the tail of the power spectrum will still be dominated by the effects of extremely weak extrinsic noise so that the $k^{-4}$ tail of the power spectrum is retained with very small $k^{-8}$ corrections (See appendix one) indicating that weak parameter noise is not a qualitatively important factor in the analysis of quasi-patterns. Absent weak extrinsic noise, there is no evidence that weak parameter noise generates quasi-patterns on its own.  Rather, it seems to only add corrections to the approach to the mean field steady state.  

A potential difficulty with the analysis of power spectrum tails is that preliminary numerical study of quasi-patterns suggests that the form of the power spectrum used above may only be strictly valid near the onset of the patterns due to higher order corrections to the mean population.  Evidence for this claim is discussed below. The implications of this possibility for distinguishing different kinds of patterns is a subject for future research.   

\section{Thermodynamic limit}

To compare to data, we also must be able to estimate the conditions under which the fluctuation driven effects described above are important.  While the considerations that follow are mathematically elementary, they are important for the analysis of real data and have not always been clearly elucidated in the ecological literature, where it has sometimes been assumed that intrinsic noise effects are only important if the total population of each species is small \cite{MAY73}.  In fact, the scale of fluctuations is governed instead by the population size in a volume (indicated by the parameter $V$ in the calculation above) sufficiently small that the time to diffuse out of the volume is smaller than the typical time between reactions per particle.  The confusion arises because when space is neglected, the organisms are all confined to such a volume, so the scale of fluctuations is determined by the total population size \cite{MCKAN04,MCKAN05}.  

The present calculation shows that there are two separate limits in the construction of reaction-diffusion models.  One of these limits yields a particular kind of mean field theory, and the other, corresponding to what would traditionally be called the thermodynamic limit in statistical physics, does not yield a mean field theory at all.  Only in $d=0$ do these limits coincide.  Recall that the theory was constructed by creating a lattice of patches, each patch of volume $V$, and then taking the limit of an infinite number of patches, and looking at the continuum version of the theory. The thermodynamic limit corresponds to the limit as the number of patches goes to infinity, while the mean field limit corresponds to taking the volume of each patch, $V$, to infinity.

The parameter $V$ can be estimated by noting that the time required on average to diffuse out of a volume $V\sim L^d$ can be estimated to be $t\sim L^2/D$ with diffusivity given by $D$.  If a characteristic reaction rate for single particles (possibly depending in complex way on concentrations) is $R$, then the size of a well mixed volume is constrained by
\begin{align}
 \frac{L^2}{D}<\frac{1}{R}
\end{align}
 \noindent Rearranging, the size of a well mixed volume is estimated by 
\begin{align}
 V\sim\left(\frac{D}{R}\right)^{d/2}
\end{align}

Multiplying the size of a locally well mixed volume by the smallest local population density will provide an estimate of how far the system is from the mean field limit.  However, care should be taken because of amplification effects as discussed in the following section.  Further information about how far the system is from the mean field limit can be obtained by examining the properties of the power spectrum or correlation functions as discussed above. 

\section{Validity of the large $V$ expansion and the scale of quasi-patterns}

\begin{figure*}[ht]
\begin{center}
\includegraphics[width=6.5in]{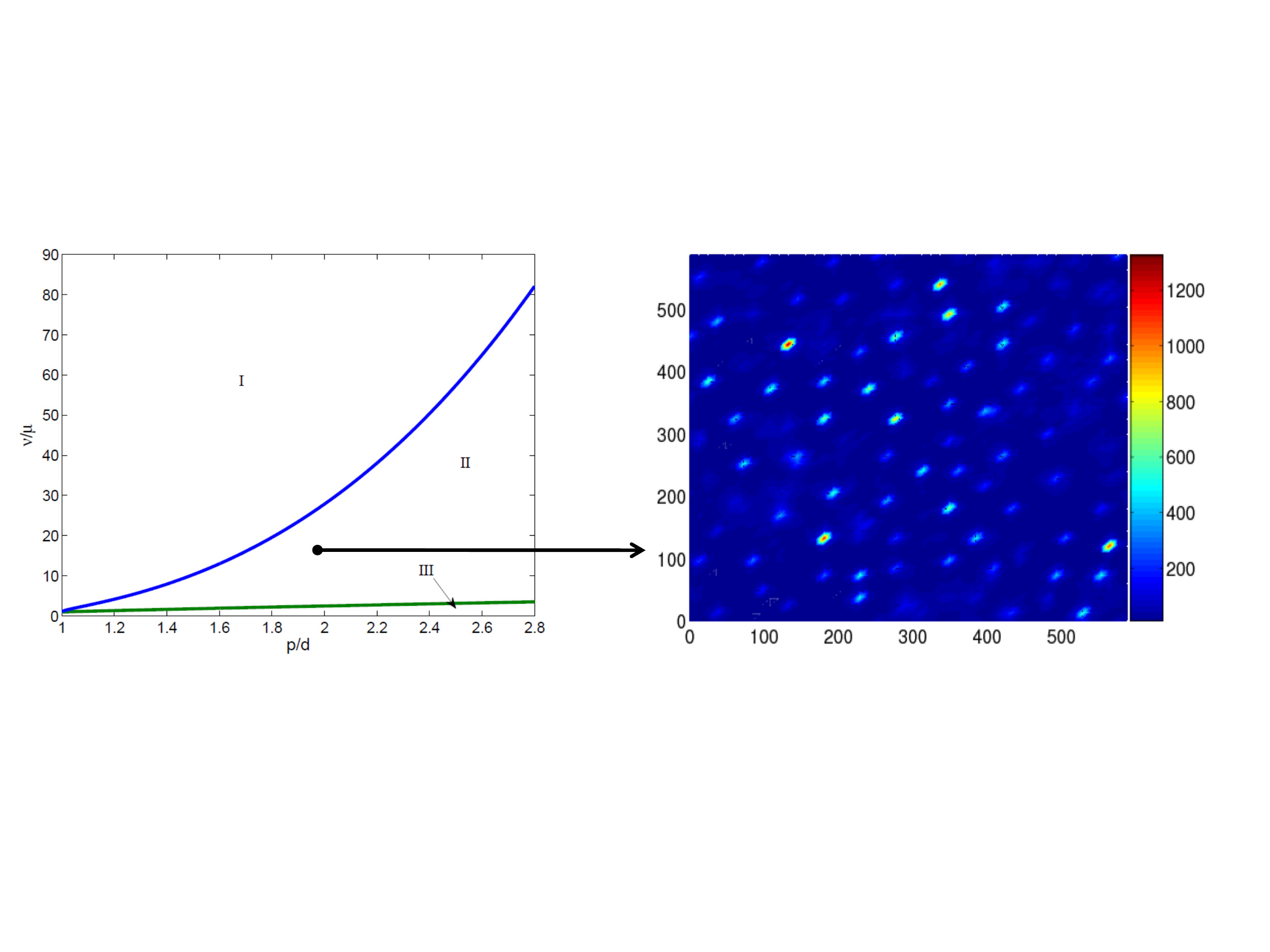}
\caption{The left hand panel is the phase diagram of the Levin-Segel model with intrinsic fluctuations.  The vertical axis is the ratio of diffusivities, and the horizontal axis is the dimensionless ratio of generic kinetic parameters.  Region I is mean field pattern formation, region II is fluctuation driven pattern formation, and region III has no spatial pattern formation.  The black arrow has its tail on the approximate location in parameter space simulated to produce the spatial patterns shown on the right.  The right hand panel is a heat map of population density in two dimensions.  Note that the number of organisms is highly variable, even though mean field predicts no spatial patterns.  The fluctuation effects are large, with patch populations ranging from 1200 to 0.  The axes are the lattice index from simulation.}
\label{ch5:fig3}
\end{center}
\end{figure*}

The expansion considered above is only strictly valid near the onset of quasi-patterns.  While in the absence of space, the expansion is valid quite generally, leading to excellent agreement between theory and simulation for the power spectrum \cite{MCKAN05}, the spatial structures do not seem to be as well captured by the expansion deep in the quasi-pattern regime.  This is probably due to fluctuation corrections to the mean field not studied in the current paper.  This is suggested by simulated trajectories of the reaction-diffusion master equation using the exact algorithm of Gillespie \cite{GILL76}.  The results of this calculation, along with the location on the phase diagram simulated are shown in fig. \ref{ch5:fig3}

The calculation indicates that the patterns deep inside the quasi-pattern phase are non-perturbative, due to the large variance in populations.  We expect that the non-perturbative corrections to the mean field solutions arise at higher order in the expansion.  The analytical theory above does not predict the power spectrum of these patterns, but the calculation of wavelength and period above are still approximately valid, since they are obtained by finding the least stable modes, which are likely still dominant, even in the non-perturbative regime.

\section{Explaining the failure of mean field theory}

From the above calculation, as well as related calculations ranging from zero dimensional models of ecosystems to models of biochemical oscillations \cite{BUTL09b,MCKAN05,MCKAN07,BLAD10} it is clear that in many applications where the fundamental physics contains intrinsic noise, mean field theory fails to describe the oscillatory dynamics in time and space of the system even for relatively large systems with many degrees of freedom far from a critical point.  Qualitatively, this failure can be understood quite generally by considering the nature of mean field theory.  

While there are many ways to derive mean field theories \cite{GOLD92}, to understand the failure of mean field theory, the simplest approach for systems described by a master equation is to note that there are two essential steps to deriving a mean field theory: averaging and neglecting correlations. 

\begin{figure}[ht]
\begin{center}
\includegraphics[width=3.5in]{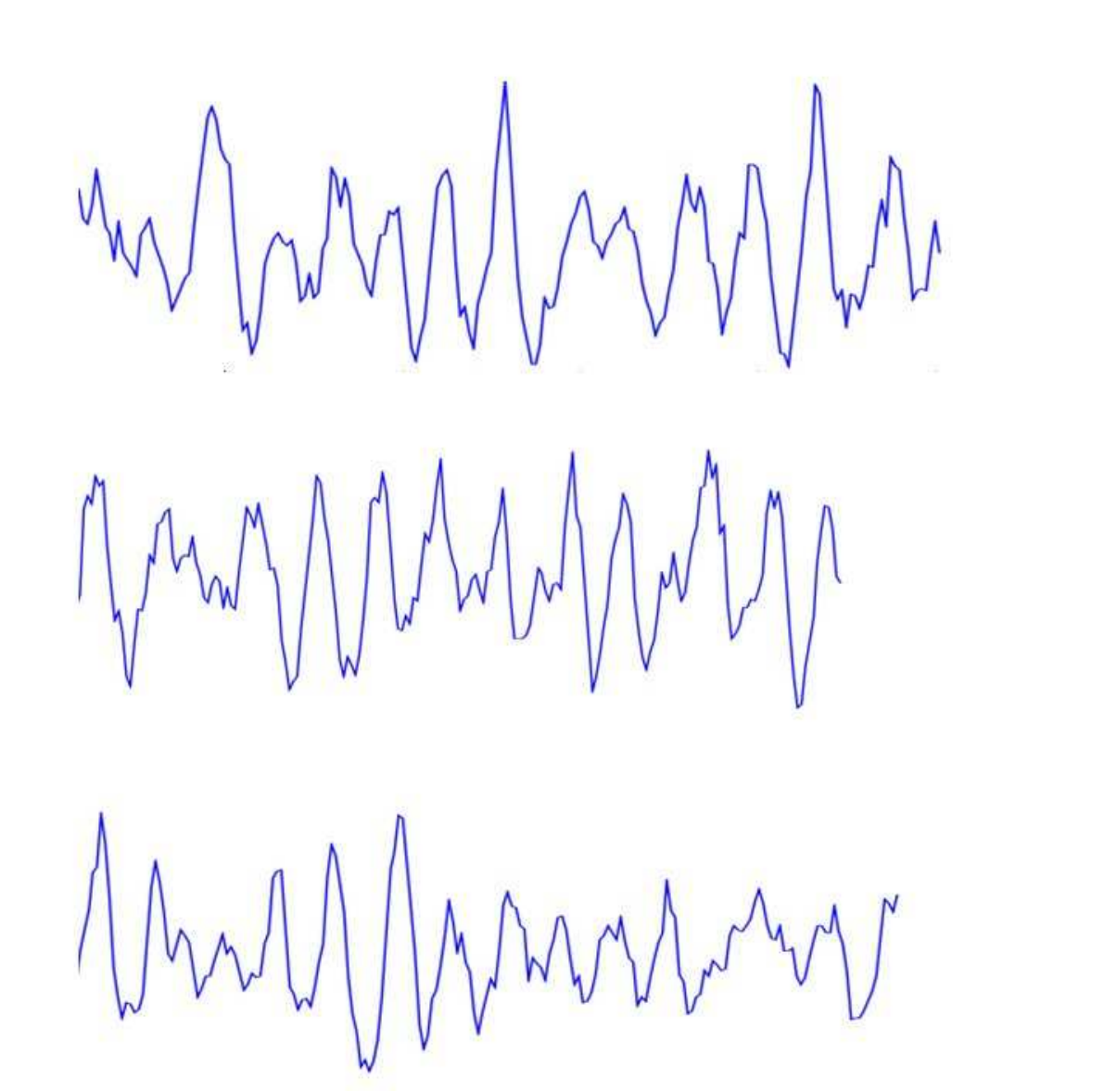}
\caption{Sample trajectories of the Markov process for predator-prey dynamics.  Note that while each is roughly oscillatory, a mean field theory derived from the average of many such trajectories would not contain oscillations.}
\label{ch5:fig4}
\end{center}
\end{figure}

Consider the first step, averaging.  The average of the trajectories is given by
\begin{align}
\varphi = \langle N(t,x) \rangle = \lim_{M_{\zeta \to \infty}} \frac{1}{M_{\zeta}}\sum_{\zeta} N_{\zeta}(t,x) 
\label{ch5:65}
\end{align}

\noindent where $\zeta$ is the index for realizations of the discrete Markov process for the population dynamics, $M_{\zeta}$ is the number of realizations sampled, and $N_{\zeta}(t,x)$ is the realization of the discrete Markov process.  Each individual realization may be oscillatory, but the oscillations will have a great deal of noise in their amplitude and phase.  Summing over these oscillatory contributions will under many conditions lead to an average, $\varphi$, that is no longer oscillatory because the variance in amplitude and phase between different realizations $\zeta$ of the stochastic process will lead to cancellations of the oscillatory parts in the sum for the average above.  That is, the sum of noisy oscillations is not always oscillatory.  Since mean field theory considers the dynamics of averages, it will not capture the oscillations present in individual realizations of the dynamics unless the feedbacks that generate the oscillations are much more important than fluctuations (see fig. \ref{ch5:fig4}).       

\section{Application to field data and experiments}

While the calculation above was intended primarily to shed light on the broad theoretical question of the fine tuning problem in Turing instabilities rather than the Levin-Segel model alone, it would still be satisfying to match the predictions above to plankton data.  Such an application to current field data in planktonic systems is very difficult.  In part this is because data on plankton patterns are primarily gathered for large scale spatial patterns that are driven by turbulent stirring, rather than biological interactions as in the theory presented here \cite{ABRA98}.  Convection accounts for most of the spatial heterogeneity of plankton at scales above tens of meters \cite{ABRA98}.  However, there do exist some limited data on plankton population heterogeneity at meter and shorter length scales \cite{DAVI92}.  Further data on the motility of plankton suggest that the ratio of diffusivities for predator-prey pairs is of order 10 \cite{VISS06}.  We calculated above that with generic parameters, the criterion \ref{ch5t:55} yields $\nu/\mu>2.48$, while the Turing condition yields $\nu/\mu>27.8$.  Under these conditions, it is likely that some populations have fluctuation driven patterns, if the Turing mechanism is responsible for the pattern formation.  Current data are not, to our knowledge, adequate to go much further.

There are several additional problems with applying the current theory to real planktonic systems, even if the data were to be much higher resolution.  The first is that plankton are enormously diverse, with many species interacting with many others, and body sizes and behaviors spanning several orders of magnitude \cite{LI02,KANE07}.  A second problem is that the current theory is so simplified that there is no clear connection between many of the parameters in the model and what is measured in real populations.  The best way to carry out the identification of quasi-patterns is probably not to engage in detailed modeling of the population dynamics, but rather to use model independent predictions, such as the density dependence of the correlations described above, and the power of $k$ and $\omega$ for large values of $k$ and $\omega$ in the power spectrum.  Data sets associated with plant systems are likely to be amenable to such analysis \cite{REIT08}.  Additionally, laboratory experiments in engineered microbial \cite{LU10}, or even in chemical systems (see above comments on the thermodynamic limit) may provide more controlled ways to detect quasi-patterns.  

\section{Conclusions and prospects for future research}

We conclude by noting that our analysis of the model in Eqs. \ref{ch5:1} has demonstrated that Turing patterns are much more generic than is to be expected on the basis of mean field theory, partial differential equation analysis.  We also have pointed out some possible ways in which the fluctuation driven spatiotemporal patterns discussed can be identified in real data.  While this paper focused on a single model, we wish to emphasize again that the model was deliberately chosen to be generic with the goal of providing broad insight into the statistical mechanics of the Turing mechanism that can be widely applied.  As noted in the introduction, the conjectured wide applicability of this result has received some support from calculations on the Brusselator model \cite{BIAN10} and a model of Turing patterns in neuronal networks studied in the following paper \cite{BUTL10}.  Further applications of this theory are potentially as wide as the applicability of the Turing mechanism, which, as was pointed out in the introduction, has been used to explain patterns in an enormous variety of systems.  In fact, we conjecture that perhaps many or most observed Turing patterns are the quasi-patterns predicted in this paper.  To demonstrate this conjecture, the next step is to apply the concepts in this paper to an experimentally well-characterized system, such as an engineered bacterial system with Turing feedbacks.  Another important avenue of investigation is to further explore ways to distinguish between quasi-patterns and mean field patterns.  Further theoretical progress may also be made by addressing with a similarly detailed theory other noise driven spatiotemporal patterns such as intrinsic noise driven epidemic waves, which seem to be present in measles and dengue fever epidemics \cite{GREN01,CUMM04}.      

\section{Acknowledgements}
This work was partially supported by National Science Foundation Grant No. NSF-EF-0526747.  Additional funding for T. B. was provided through a Drickamer fellowship through the University of Illinois department of physics.  

\appendix
\section{Influence of parameter noise on quasi-patterns}
To calculate the effects of parameter noise on Turing models to a first approximation, we take the linearized mean field dynamics
\begin{equation}
-i\omega \bm{x}=\bm{A}\bm{x}
\label{A1}
\end{equation}
\noindent and generalize to 
\begin{equation}
-i\omega \bm{x}=\bm{A}\bm{x}+\epsilon \xi\bm{x} +\bm{x_0}
\label{A3}
\end{equation}
\noindent where $\bm{x_0}$ is an initial deviation from mean field equilibrium, $\epsilon$ is a small parameter, and $\xi$ is a diagonal matrix of white noise, variance one.  Only contributions to the power spectrum that are independent of $\bf{x_0}$ persist in the long time limit.  

Rearranging Eq. \ref{A3} yields
\begin{align}
\bm{x}=-\left(\bm{A}-\epsilon \xi_1\right)^{-1}\bm{x}_0 
\label{A4}
\end{align}
For small $\epsilon$, this can be expanded to yield
\begin{equation}
 \bm{x}=\bm{A}^{-1}\left(\bm{1}+\epsilon\bm{A}^{-1}\xi\right)\bm{x}_0+ O(\epsilon^2)
\label{A5}
\end{equation}
Taking the dot product and averaging yields the sum of the power spectrum for the predator and prey species, which is sufficient for detecting the presence or absense of quasi-patterns

\begin{equation}
 \langle \bm{x} \bm{x}^* \rangle =  \left(\bm{A}^{-1}\bm{x}_0\right)\left(\bm{A}^{-1}\bm{x}_0\right)^* + \epsilon^2\bm{A}^{-2}\bm{x}_0 (\bm{A}^{-2}\bm{x}_0)^*
\label{A6}
\end{equation}

Note that this power spectrum depends term by term on the initial conditions $\bm{x}_0$, indicating that the power spectrum is dominated completely by the effects of transient patterns at times shorter than the relaxation time to the steady state.  Parameter noise only has the effect of correcting the approach to steady state. Any qualitatively important effects of parameter noise on quasi-patterns are thus present only in a nonlinear analysis. 

\bibliographystyle{apsrev}
\bibliography{thesisbib}

\end{document}